\documentclass[a4paper, amsfonts, amssymb, amsmath, reprint, showkeys, nofootinbib, onecolumn]{revtex4-1}
\usepackage[english]{babel}
\usepackage{graphicx}% Include figure files
\usepackage{dcolumn}% Align table columns on decimal point
\usepackage{bm}% bold math
\usepackage{subcaption}
\usepackage{comment}
\usepackage{xcolor}
\usepackage{float}
\usepackage{titlesec}
\usepackage[pdftex, pdftitle={Article}, pdfauthor={Author}]{hyperref}
\definecolor{purple}{rgb}{0.58,0.0,0.83}

\definecolor{blue(pigment)}{rgb}{0.2, 0.2, 0.6}

\usepackage{scalerel}
\usepackage{tikz}
\usetikzlibrary{svg.path}

\definecolor{orcidlogocol}{HTML}{A6CE39}
\tikzset{
  orcidlogo/.pic={
    \fill[orcidlogocol] svg{M256,128c0,70.7-57.3,128-128,128C57.3,256,0,198.7,0,128C0,57.3,57.3,0,128,0C198.7,0,256,57.3,256,128z};
    \fill[white] svg{M86.3,186.2H70.9V79.1h15.4v48.4V186.2z}
                 svg{M108.9,79.1h41.6c39.6,0,57,28.3,57,53.6c0,27.5-21.5,53.6-56.8,53.6h-41.8V79.1z M124.3,172.4h24.5c34.9,0,42.9-26.5,42.9-39.7c0-21.5-13.7-39.7-43.7-39.7h-23.7V172.4z}
                 svg{M88.7,56.8c0,5.5-4.5,10.1-10.1,10.1c-5.6,0-10.1-4.6-10.1-10.1c0-5.6,4.5-10.1,10.1-10.1C84.2,46.7,88.7,51.3,88.7,56.8z};
  }
}

\newcommand\orcidicon[1]{\href{https://orcid.org/#1}{\mbox{\scalerel*{
\begin{tikzpicture}[yscale=-1,transform shape]
\pic{orcidlogo};
\end{tikzpicture}
}{|}}}}

\begin{document}

\title{Thermodynamic Criticality in FLRW Cosmology with Non-extensive Loop Quantum Gravity Entropy}

\author{Miguel Cruz$^1$\orcidicon{0000-0003-3826-1321}}
\email{miguelcruz02@uv.mx}

\author{Joel Saavedra$^2$\orcidicon{0000-0002-1430-3008}}
\email{joel.saavedra@pucv.cl}

\author{Juan Rodriguez$^2$}
\email{juan.rodriguez.f@mail.pucv.cl}

\affiliation{$^1$Facultad de F\'{\i}sica, Universidad Veracruzana 91097, Xalapa, Veracruz, M\'exico.\\
$^2$Instituto de F\'\i sica, Pontificia Universidad Cat\'olica de Valpara\'\i so, Casilla 4950, Valpara\'\i so, Chile.}

\begin{abstract}
We investigate thermodynamic criticality in a spatially flat Friedmann-Lema\^{\i}tre-Robertson-Walker universe with a non-extensive Loop Quantum Gravity inspired entropy on its apparent horizon. Using the full Kodama-Hayward temperature and the unified first law, we derive the modified Friedmann dynamics and construct the corresponding horizon equation of state. We see that a finite physical critical point only appears in the non-extensive branch $q>1$, while in the Bekenstein-Hawking limit $q\to1$, the critical point is continuously pushed to $v_c\to\infty$ and $T_c\to0$. Along the same spinodal curve, both the constant-pressure heat capacity and the isothermal compressibility become unbounded, and the extremum of this curve is located at the critical point. Below the critical temperature, the Gibbs free energy develops several thermodynamic branches, and the phase coexistence for different horizon states is verified by the equality of temperature, pressure, and Gibbs free energy. The critical exponents are $(\alpha_{cr},\beta_{cr},\gamma_{cr},\delta_{cr}) =(0,1/2,1,3)$, indicating that the system belongs to the standard mean-field universality class despite the non-algebraic form of the equation of state. The normalized Ruppeiner curvature diverges precisely on the spinodal curve and shows critical scaling $R_N\sim-|v-v_c|^{-4}$ on the critical isotherm, and $R_N\sim-|t|^{-2}$ on the critical isochore. Finally, the critical expansion scale is given by $H_c^2=2(\sqrt{5}-2)|\beta|$, which implies that the thermodynamic critical point occurs when the entropy deformation is of order unity. These results establish a self-consistent critical structure for the effective thermodynamic state space of the cosmological apparent horizon and retain the mean-field critical universality.
\end{abstract}

\maketitle
\section{Introduction}
\label{sec:introduction}
Among the most suggestive hints of microscopic degrees of freedom of spacetime are those arising from the relationship between gravitation and thermodynamics. The laws of black hole mechanics, the Bekenstein-Hawking area law and Hawking radiation, provided a consistent assignment of entropy and temperature to gravitational horizons. Subsequently, this relation was generalized to the dynamics of spacetime itself, including the derivation of gravitational field equations from thermodynamic relations \cite{Bekenstein1973,BardeenCarterHawking1973,Hawking:1975,GibbonsHawking1977,Jacobson:1995,Padmanabhan2010}. In cosmological spacetimes, the apparent horizon provides a natural boundary to define thermodynamic quantities. Hayward's unified first law is a thermodynamic description of a Friedmann-Lema\^{\i}tre-Robertson-Walker (FLRW) universe. In this formulation, the matter sector is included by means of the work density and the energy-supply vector, and the Kodama-Hayward temperature is determined by the dynamical surface gravity \cite{Hayward1998,BakRey2000,CaiKim2005,AkbarCai2007,CaiCaoHu2008}.
The construction does not require the apparent horizon to be stationary. In particular, keeping the full Kodama-Hayward temperature retains the contribution related to the dynamical evolution of the horizon and is therefore suitable for the study of truly dynamical thermodynamic phenomena.

A major generalization of this framework is the replacement of the Bekenstein-Hawking entropy by a generalized entropy.
The motivation for these modifications comes from non-extensive statistical mechanics, quantum-gravity considerations, fractal horizon structures, and corrections to the microscopic counting of horizon states. As representative examples, we mention the Tsallis, R\'enyi, Barrow, Kaniadakis, and Loop Quantum Gravity motivated entropies \cite{Tsallis1988,CzinnerIguchi2016,Barrow2020,Kaniadakis2002,LymperisSaridakis2018,LymperisBasilakosSaridakis2021,Saridakis2020,NojiriOdintsovFaraoni2022,NojiriOdintsovPaul2022}.
Ascribing such entropy functionals to the cosmological apparent horizon modifies the thermodynamic derivation of the Friedmann equations, leading to effective cosmological dynamics that differ from those of standard General Relativity.

Within this more general framework, Majhi showed that non-extensive statistical mechanics can be consistently embedded in the microscopic Loop Quantum Gravity formulation of black-hole entropy \cite{Majhi2017}. On the basis of this construction, Lymperis wrote down the corresponding LQG-inspired entropy in the exponential non-extensive form that is used in the present work and studied the implications for cosmology \cite{Lymperis2023}. This entropy thus provides a concrete arena in which a quantum-geometry-inspired change of horizon microphysics can be propagated into the macroscopic dynamics of an expanding universe.

A complementary development is that of thermodynamic phase transitions in gravitational systems. Charged AdS black holes have Van der Waals-like isotherms, first-order phase transitions, spinodal instabilities, divergent response functions, and mean-field critical exponents in extended black hole thermodynamics \cite{KubiznakMann2012,KubiznakMannTeo2017}. In more recent times, similar constructions have been developed for cosmological apparent horizons \cite{Kong2022,Abdusattar2022,Kong2023,AbdusattarScalarTensor2023,Housset2024,Rivadeneira2026,GonzalezEspinoza2026,CruzFractional2026}. In this context, the thermodynamic pressure is identified with the work density, and the specific volume is taken to be proportional to the apparent-horizon radius.
A particularly important result is that the usual Einstein-FLRW thermodynamics does not have a finite $P$-$V$ critical point \cite{Kong2022}. However, nontrivial critical configurations can be produced by modified theories of gravity or generalized horizon entropies \cite{AbdusattarScalarTensor2023,Housset2024,Rivadeneira2026}.
This leads us to the main question, which is the subject of the present work:
can the non-extensive LQG-inspired entropy give rise to real thermodynamic criticality on the cosmological apparent horizon while the Bekenstein-Hawking system has no finite critical point?
This question is nontrivial in the sense that the LQG-inspired entropy is exponentially dependent on the area of the horizon. The corresponding Friedmann dynamics and horizon equation of state are thus non-algebraic, with exponential-integral contributions. This entropy has been previously considered in terms of the cosmological dynamics \cite{Lymperis2023}, but not, to our knowledge, systematically investigated in terms of its full thermodynamic phase structure at the apparent horizon.

In this work we tackle this problem by means of the full dynamical Kodama-Hayward formalism. We obtain the modified Friedmann equations from the unified first law without using a quasi-static or adiabatic approximation. Then we identify the thermodynamic pressure with the work density and introduce the specific volume $v=2R_A$ to formulate the exact horizon equation of state.
We analytically demonstrate that a finite positive-temperature critical point only exists in the non-extensive branch $q>1$. In the Bekenstein-Hawking limit, the critical point is moved to the infinite specific volume and zero critical temperature, which is consistent with the absence of finite criticality in the standard Einstein-FLRW thermodynamics \cite{Kong2022}. The heat capacities, the isothermal compressibility, the spinodal structure, and the Gibbs free energy are analyzed, and phase coexistence between different horizon thermodynamic states is established. It is shown that the critical exponents are the usual mean-field values $(\alpha_{cr},\beta_{cr},\gamma_{cr},\delta_{cr})=(0,1/2,1,3)$.
We further investigate the thermodynamic geometry by means of the normalized Ruppeiner curvature. Its singular locus coincides with the spinodal instability, and its critical divergence is directly related to the divergence of the isothermal compressibility. The combined analysis reveals a clear separation between global and local critical properties: the entropy deformation controls the existence and location of the critical point, while the local analytic structure of the equation of state retains the standard mean-field universality class.

The paper is structured as follows. In Sec.~\ref{sec:entropy} we introduce the non-extensive LQG-inspired entropy on the cosmological apparent horizon. In Sec.~\ref{sec:friedmann} we obtain the modified Friedmann equations from the unified first law. The horizon equation of state and its Bekenstein-Hawking limit are discussed in Sec.~\ref{sec:eos}. We then determine the critical point and study the response functions, spinodal structure, Gibbs free energy and critical exponents. The cosmological significance of the critical scale is discussed in Sec.~\ref{sec:cosmological} and the thermodynamic geometry is explored in Sec.~\ref{sec:ruppeiner}. We conclude by summarizing our results and discussing their physical implications.

\section{Non-extensive Loop Quantum Gravity Inspired Entropy} \label{sec:entropy}
We apply the non-extensive entropy proposed by Majhi and then applied to cosmology by Lymperis, motivated by Loop Quantum Gravity, to include the quantum-geometric corrections in the thermodynamics of the cosmological apparent horizon. The construction unites the microscopic counting of horizon states in Loop Quantum Gravity \cite{Rovelli1996,AshtekarBaezCorichiKrasnov1998} with a non-extensive statistical description, which in turn entails an exponential deformation of the Bekenstein-Hawking area law.
The entropy is given by
\begin{equation}
S_{\mathrm{LQG}}
=
\frac{1}{1-q}
\left[
e^{(1-q)\Lambda S_{\mathrm{BH}}}-1
\right],
\label{eq:SLQG}
\end{equation}
where $S_{\mathrm{BH}}=A/4G$ is the Bekenstein-Hawking entropy, $q$ is the non-extensive entropic
index, and $\Lambda=\Lambda(\gamma_0)$ is a dimensionless quantity
determined by the Barbero-Immirzi parameter $\gamma_0$. The value of this parameter is fixed by the microscopic counting of quantum-geometry horizon states, which reproduces the Bekenstein-Hawking area law \cite{Rovelli1996,AshtekarBaezCorichiKrasnov1998}. In the
underlying Loop Quantum Gravity construction,
\begin{equation}
\Lambda(\gamma_0)
=
\frac{\ln 2}{\sqrt{3}\pi\gamma_0}.
\label{eq:Lambda}
\end{equation}
For the canonical choice $\gamma_0=\ln 2/(\pi\sqrt{3})$, one obtains $\Lambda=1$. In the extensive limit $q\rightarrow1$, Eq.~\eqref{eq:SLQG} reduces
to
\begin{equation}
\lim_{q\rightarrow1}S_{\mathrm{LQG}}
=
\Lambda S_{\mathrm{BH}}.
\label{eq:q1limit}
\end{equation}
Thus, for the canonical normalization $\Lambda=1$, the standard
Bekenstein-Hawking area law is recovered. For a spatially flat FLRW universe, the apparent-horizon radius is $R_A=1/H$ and its area is $A=4\pi R_A^2$. Consequently,
\begin{equation}
S_{\mathrm{BH}}
=
\frac{\pi R_A^2}{G},
\label{eq:SBH_RA}
\end{equation}
and it is convenient to introduce
\begin{equation}
\beta
=
\frac{\pi(1-q)\Lambda}{G}.
\label{eq:beta}
\end{equation}
The entropy then takes the compact form $S_{\mathrm{LQG}}=(1-q)^{-1}\left(e^{\beta R_A^2}-1\right)$. For $G>0$ and $\Lambda>0$, the sign of $\beta$ is determined entirely by the non-extensive parameter $q$, namely $\beta<0$ if and only if $q>1$. This condition is not simply a convenient choice of parameters, as we show below. This is a direct consequence of the requirement that the critical specific volume be real and positive. Thus, the non-extensive entropy $q>1$ branch is selected by thermodynamic criticality. This compact form of the entropy provides the thermodynamic input for the cosmological construction in the following sections. The modification of Friedmann dynamics due to the unified first law is determined by its derivative in relation to the apparent-horizon radius.

\section{Modified Friedmann Equations} \label{sec:friedmann}
Using the unified first law at the apparent horizon, we now obtain the cosmological dynamics induced by the non-extensive Loop Quantum Gravity motivated entropy. For a spatially flat FLRW universe, the entropy introduced in Sec.~\ref{sec:entropy} can be written as
\begin{equation}
S_{\mathrm{LQG}}
=
\frac{1}{1-q}
\left(
e^{\beta R_A^2}-1
\right),
\qquad
\beta=
\frac{\pi(1-q)\Lambda}{G}.
\end{equation}
Differentiating with respect to the apparent-horizon radius gives
\begin{equation}
dS_{\mathrm{LQG}}
=
\frac{2\pi\Lambda R_A}{G}
e^{\beta R_A^2}
\,dR_A.
\label{eq:dSLQG}
\end{equation}
The gravitational field equations can be written in the unified
first-law form \cite{Hayward1998,CaiKim2005,AkbarCai2007}. The unified first law projected along the apparent horizon is written
in Clausius form as
\begin{equation}
\delta Q
=
-dE+W\,dV
=
T\,dS_{\mathrm{LQG}},
\label{eq:firstlaw}
\end{equation}
where $W=\tfrac{1}{2}(\rho-p)$ is the work density,
\begin{equation}
V=\frac{4\pi}{3}R_A^3
\label{eq:volume}
\end{equation}
is the volume enclosed by the apparent horizon, and
\begin{equation}
E=\rho V
=
\frac{4\pi}{3}\rho R_A^3
\label{eq:MSenergy}
\end{equation}
is the Misner-Sharp energy. The energy enclosed by the apparent horizon is identified with the
Misner-Sharp energy \cite{MisnerSharp1964, Hayward1998}. Taking the time derivative of Eq.~\eqref{eq:MSenergy} yields
\begin{equation}
\dot E
=
4\pi R_A^2\rho\,\dot R_A
+
\frac{4\pi}{3}R_A^3\dot\rho ,
\end{equation}
while
\begin{equation}
\dot V=4\pi R_A^2\dot R_A .
\end{equation}
Therefore,
\begin{equation}
-\dot E+W\dot V
=
-2\pi R_A^2(\rho+p)\dot R_A
-\frac{4\pi}{3}R_A^3\dot\rho .
\end{equation}
Using the continuity equation
\begin{equation}
\dot\rho+3H(\rho+p)=0,
\label{eq:continuity}
\end{equation}
the heat flow can be expressed as
\begin{equation}
-\dot E+W\dot V
=
4\pi H R_A^3(\rho+p)
\left(
1-\frac{\dot R_A}{2HR_A}
\right).
\label{eq:heatflow}
\end{equation}
For a dynamical apparent horizon, the appropriate temperature is
associated with the Kodama-Hayward surface gravity
\cite{Kodama1980,Hayward1998},
\begin{equation}
\kappa
=
-\frac{1}{R_A}
\left(
1-\frac{\dot R_A}{2HR_A}
\right).
\label{eq:kappa}
\end{equation}
We adopt the corresponding positive horizon temperature
\begin{equation}
T
=
-\frac{\kappa}{2\pi}
=
\frac{1}{2\pi R_A}
\left(
1-\frac{\dot R_A}{2HR_A}
\right).
\label{eq:KHtemperature}
\end{equation}
No quasi-static approximation is imposed in
Eq.~\eqref{eq:KHtemperature}. From Eq.~\eqref{eq:dSLQG},
\begin{equation}
\dot S_{\mathrm{LQG}}
=
\frac{2\pi\Lambda R_A}{G}
e^{\beta R_A^2}
\dot R_A ,
\end{equation}
and therefore
\begin{equation}
T\dot S_{\mathrm{LQG}}
=
\frac{\Lambda}{G}
e^{\beta R_A^2}
\dot R_A
\left(
1-\frac{\dot R_A}{2HR_A}
\right).
\label{eq:TdotS}
\end{equation}
Substitution of Eqs.~\eqref{eq:heatflow} and
\eqref{eq:TdotS} into Eq.~\eqref{eq:firstlaw} shows that the full
dynamical factor appearing in the Kodama-Hayward temperature occurs
identically on both sides of the Clausius relation and therefore
cancels exactly. One obtains
\begin{equation}
\frac{\Lambda}{G}
e^{\beta R_A^2}\dot R_A
=
4\pi H R_A^3(\rho+p).
\label{eq:basicdyn}
\end{equation}
Using Eq.~\eqref{eq:continuity}, this relation becomes
\begin{equation}
\dot\rho
=
-\frac{3\Lambda}{4\pi G}
\frac{e^{\beta R_A^2}}{R_A^3}
\dot R_A ,
\end{equation}
or equivalently
\begin{equation}
d\rho
=
-\frac{3\Lambda}{4\pi G}
\frac{e^{\beta R_A^2}}{R_A^3}
\,dR_A.
\label{eq:drho}
\end{equation}
Integrating Eq.~\eqref{eq:drho} gives
\begin{equation}
\rho(R_A)
=
\frac{3\Lambda}{8\pi G}
\left[
\frac{e^{\beta R_A^2}}{R_A^2}
-
\beta\operatorname{Ei}
\left(
\beta R_A^2
\right)
\right]
+\rho_0 ,
\label{eq:rhoRA}
\end{equation}
where $\rho_0$ is an integration constant. In the following we set
$\rho_0=0$, since a nonzero value can be absorbed into an independent
vacuum-energy contribution. Using $R_A=1/H$, Eq.~\eqref{eq:rhoRA} becomes
\begin{equation}
\rho(H)
=
\frac{3\Lambda}{8\pi G}
\left[
H^2e^{\beta/H^2}
-
\beta\operatorname{Ei}
\left(
\frac{\beta}{H^2}
\right)
\right].
\label{eq:FriedmannLQG}
\end{equation}

Equation~\eqref{eq:FriedmannLQG} is the modified first Friedmann
equation generated by the non-extensive entropy. Its non-algebraic
structure follows directly from the exponential dependence of the
entropy on the apparent-horizon area.
As a consistency check, consider the extensive limit $q\rightarrow1$, $\beta\rightarrow0$. Since $\beta\operatorname{Ei}\left(\beta/H^2\right)\longrightarrow0$, Eq.~\eqref{eq:FriedmannLQG} reduces to
\begin{equation}
\rho
=
\frac{3\Lambda}{8\pi G}H^2.
\end{equation}
For the canonical normalization $\Lambda=1$, one therefore recovers
the standard Friedmann equation,
\begin{equation}
H^2
=
\frac{8\pi G}{3}\rho .
\label{eq:FriedmannGR}
\end{equation}

\section{Equation of State}
\label{sec:eos}

We now construct the thermodynamic equation of state associated with
the cosmological apparent horizon. Following the thermodynamic formulation of the FLRW apparent horizon,
we identify the work density with the thermodynamic pressure,
\begin{equation}
P=W=\frac{\rho-p}{2},
\end{equation}
as considered in Refs.~\cite{Kong2022,Abdusattar2022}.
Since
\begin{equation}
P
=
\rho-\frac{1}{2}(\rho+p),
\label{eq:P_rhop}
\end{equation}
the equation of state can be obtained once the combination
$\rho+p$ is expressed in terms of the horizon temperature and radius. From Eq.~\eqref{eq:basicdyn},
\begin{equation}
\frac{\Lambda}{G}
e^{\beta R_A^2}\dot R_A
=
4\pi H R_A^3(\rho+p).
\end{equation}
The Kodama-Hayward temperature,
\begin{equation}
T
=
\frac{1}{2\pi R_A}
\left(
1-\frac{\dot R_A}{2HR_A}
\right),
\end{equation}
gives
\begin{equation}
\dot R_A
=
2HR_A
\left(
1-2\pi R_A T
\right).
\end{equation}
Consequently,
\begin{equation}
\rho+p
=
\frac{\Lambda}{2\pi G R_A^2}
e^{\beta R_A^2}
\left(
1-2\pi R_A T
\right).
\label{eq:rhop}
\end{equation}
Using Eq.~\eqref{eq:rhoRA} with the integration constant fixed to
$\rho_0=0$, Eqs.~\eqref{eq:P_rhop} and \eqref{eq:rhop} yield
\begin{equation}
P(T,R_A)
=
\frac{\Lambda T}{2G R_A}
e^{\beta R_A^2}
+
\frac{\Lambda}{8\pi G R_A^2}
e^{\beta R_A^2}
-
\frac{3\Lambda\beta}{8\pi G}
\operatorname{Ei}
\left(
\beta R_A^2
\right).
\label{eq:EOS_RA}
\end{equation}

\begin{figure}[t]
\centering
\includegraphics[width=0.48\textwidth]{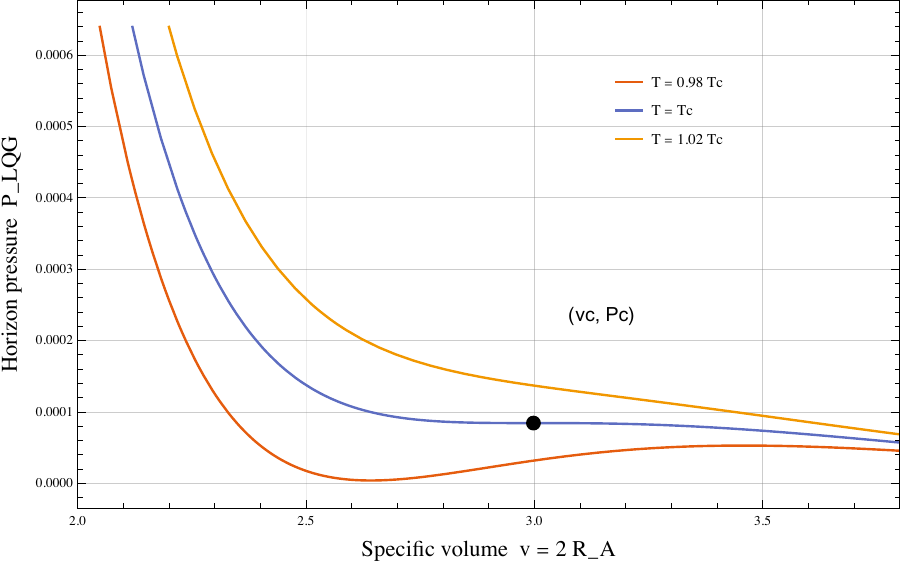}
\caption{Isotherms of the LQG-FLRW equation of state for temperatures
$T/T_c=0.98$, $1.00$, and $1.02$. At $T=T_c$, the point
$(v_c,P_c)$ is an inflection point of the critical isotherm. For
$T<T_c$, a local maximum and minimum appear, defining multiple
thermodynamic branches separated by a mechanically unstable region.
For $T>T_c$, the equation of state becomes monotonic.}
\label{fig:isotherms}
\end{figure}

To facilitate comparison with fluid and black-hole thermodynamics,
we introduce the specific volume $v=2R_A$ and define
\begin{equation}
x=\frac{\beta v^2}{4}.
\label{eq:x}
\end{equation}
The exact equation of state then becomes
\begin{equation}
P(T,v)
=
\frac{\Lambda}{G}e^x
\left(
\frac{T}{v}
+
\frac{1}{2\pi v^2}
\right)
-
\frac{3\Lambda\beta}{8\pi G}
\operatorname{Ei}(x).
\label{eq:EOS}
\end{equation}
The central thermodynamic relation of the model is given by Eq.~\eqref{eq:EOS}. The temperature dependent part is modulated by the exponential entropy deformation, while the exponential-integral term is induced by the modified Friedmann dynamics. Their combination yields a non-algebraic equation of state, which exhibits a phase structure qualitatively different from the standard FLRW apparent horizon. The Bekenstein-Hawking limit provides an immediate consistency check. For $q\rightarrow1$, $\Lambda\rightarrow1$ and $\beta\rightarrow0$, one has $e^x\rightarrow1$ and $\beta\operatorname{Ei}(x)\rightarrow0$, so that Eq.~\eqref{eq:EOS} reduces to
\begin{equation}
P(T,v)
=
\frac{T}{v}
+
\frac{1}{2\pi v^2}.
\label{eq:EOS_GR}
\end{equation}
This is the standard equation of state of the apparent horizon in a
spatially flat FLRW universe. As shown in the following section,
Eq.~\eqref{eq:EOS_GR} does not possess a finite critical point,
whereas the non-extensive equation of state
Eq.~\eqref{eq:EOS} does so for an appropriate branch of the
deformation parameter. This result is consistent with the absence of a finite critical point
in the standard Einstein-FLRW thermodynamic description
\cite{Kong2022}. For completeness, retaining the integration constant $\rho_0$ in
Eq.~\eqref{eq:rhoRA} would add a constant term to the pressure,
\begin{equation}
P(T,v)\rightarrow P(T,v)+\rho_0.
\end{equation}
Such a term does not affect the criticality conditions involving derivatives with respect to $v$ and thus leaves $v_c$ and $T_c$ unchanged. However, it shifts the critical pressure.
For the rest of this work, we adopt the vacuum-energy normalization of the thermodynamic pressure by setting $\rho_0=0$. The representative isotherms of Eq.~\eqref{eq:EOS} around the critical temperature are shown in Fig.~\ref{fig:isotherms}. For $T<T_c$, a local maximum and a local minimum appear, i.e., there are multiple thermodynamic branches separated by a mechanically unstable region.
At $T=T_c$ the two extrema merge into one inflection point and for $T>T_c$ the isotherms are monotonic.
The qualitative structure of the equation of state is illustrated in
Fig.~\ref{fig:isotherms}, where the emergence of a characteristic
Van der Waals-like oscillatory behavior below the critical temperature
can already be identified.

\section{Critical Point}
\label{sec:critical}

The critical point is determined by the standard inflection-point
conditions on the isotherm
\cite{KubiznakMann2012,Stanley1971}.
\begin{equation}
\left(\frac{\partial P}{\partial v}\right)_T=0,
\qquad
\left(\frac{\partial^2P}{\partial v^2}\right)_T=0.
\label{eq:criticalconditions}
\end{equation}
For the equation of state~\eqref{eq:EOS}, the first derivative is
\begin{equation}
\left(\frac{\partial P}{\partial v}\right)_T
=
\frac{\Lambda e^x}{2\pi Gv^3}
\left(
\pi T\beta v^3
-2\pi Tv
-\beta v^2
-2
\right),
\label{eq:Pv}
\end{equation}
where $x=\beta v^2/4$, as defined in Eq.~\eqref{eq:x}. The second derivative is
\begin{equation}
\left(\frac{\partial^2P}{\partial v^2}\right)_T
=
\frac{\Lambda e^x}{4\pi Gv^4}
\left(
\pi T\beta^2v^5
-2\pi T\beta v^3
+8\pi Tv
-\beta^2v^4
+12
\right).
\label{eq:Pvv}
\end{equation}
Eliminating the temperature between Eqs.~\eqref{eq:Pv} and
\eqref{eq:Pvv}, and defining
\begin{equation}
y=\beta v_c^2,
\end{equation}
one obtains
\begin{equation}
y^2+8y-4=0.
\label{eq:ycritical}
\end{equation}
The two algebraic roots are
\begin{equation}
y_{\pm}
=
-4\pm2\sqrt5.
\label{eq:yroots}
\end{equation}
The temperature obtained from Eq.~\eqref{eq:Pv} is
\begin{equation}
T_c
=
\frac{y+2}{\pi v_c(y-2)}.
\label{eq:Tcy}
\end{equation}
For the root
\begin{equation}
y_+
=
-4+2\sqrt5>0,
\end{equation}
the requirement $v_c^2>0$ implies $\beta>0$, but
Eq.~\eqref{eq:Tcy} then gives $T_c<0$. Therefore this branch does not
correspond to a physical positive-temperature critical point. The physical branch is instead $y_c=-4-2\sqrt5<0$, for which $v_c^2>0$ requires $\beta<0$. Using $\beta=\pi(1-q)\Lambda/G$, with $G>0$ and $\Lambda>0$, this condition is equivalent to $q>1$. Hence, a positive-temperature critical point exists only in the
non-extensive branch $q>1$. The critical specific volume is
\begin{equation}
v_c^2
=
\frac{-4-2\sqrt5}{\beta},
\label{eq:vc}
\end{equation}
while the critical temperature can be written in the compact form
\begin{equation}
T_c
=
\frac{\sqrt5-1}{2\pi v_c}.
\label{eq:Tc}
\end{equation}
At the critical point,
\begin{equation}
x_c
=
\frac{\beta v_c^2}{4}
=
-1-\frac{\sqrt5}{2},
\label{eq:xc}
\end{equation}
and the critical pressure follows directly from
Eq.~\eqref{eq:EOS},
\begin{equation}
P_c
=
\frac{\Lambda}{Gv_c^2}
\left[
\frac{\sqrt5}{2\pi}
e^{-1-\sqrt5/2}
+
\frac{3(2+\sqrt5)}{4\pi}
\operatorname{Ei}
\left(
-1-\frac{\sqrt5}{2}
\right)
\right].
\label{eq:Pc}
\end{equation}
For the representative choice $q=1.3$, $\Lambda=1$ and $G=1$, one obtains
\begin{equation}
v_c\simeq2.99820,
\qquad
T_c\simeq0.065615,
\qquad
P_c\simeq8.4013\times10^{-5}.
\label{eq:criticalnumbers}
\end{equation}
For the canonical choice $\Lambda=G=1$, the dimensionless critical ratio takes the numerical value
\begin{equation}
Z_c
=
\frac{P_cv_c}{T_c}
\simeq
3.8389\times10^{-3}.
\label{eq:Zc}
\end{equation}
The critical isotherm develops an inflection point at
$(v_c,P_c)$. For $T<T_c$, the equation of state possesses two
stationary points and therefore multiple thermodynamic branches,
whereas for $T>T_c$ the isotherms are monotonic.

\section{Thermodynamic Response Functions}
\label{sec:response}
Thermodynamic response functions provide information about the local
behavior of the horizon system and become singular at the boundaries
of thermodynamic stability. We consider the heat capacities at
constant volume and constant pressure, together with the isothermal
compressibility. The heat capacity at constant volume is defined as
\begin{equation}
C_V
=
T
\left(
\frac{\partial S}{\partial T}
\right)_V .
\label{eq:CVdef}
\end{equation}
Since
\begin{equation}
V=\frac{\pi v^3}{6},
\end{equation}
fixing the thermodynamic volume fixes $v$ and therefore the
apparent-horizon radius. Because the entropy depends only on the
horizon radius,
\begin{equation}
S_{\mathrm{LQG}}
=
\frac{1}{1-q}
\left(
e^{\beta v^2/4}-1
\right),
\end{equation}
it follows immediately that $C_V=0$. This result will imply the critical exponent
$\alpha_{cr}=0$. The heat capacity at constant pressure is
\begin{equation}
C_P
=
T
\left(
\frac{\partial S}{\partial T}
\right)_P.
\label{eq:CPdef}
\end{equation}
At fixed pressure,
\begin{equation}
dP
=
\left(\frac{\partial P}{\partial T}\right)_v dT
+
\left(\frac{\partial P}{\partial v}\right)_T dv
=
0,
\end{equation}
and hence
\begin{equation}
\left(\frac{dT}{dv}\right)_P
=
-
\frac{
\left(\frac{\partial P}{\partial v}\right)_T
}{
\left(\frac{\partial P}{\partial T}\right)_v
}.
\end{equation}
Using
\begin{equation}
\frac{dS_{\mathrm{LQG}}}{dv}
=
\frac{\pi\Lambda v}{2G}e^x,
\qquad
\left(\frac{\partial P}{\partial T}\right)_v
=
\frac{\Lambda}{Gv}e^x,
\end{equation}
together with Eq.~\eqref{eq:Pv}, one obtains
\begin{equation}
C_P
=
-
\frac{
\pi^2\Lambda T v^3 e^x
}{
G\left(
\pi T\beta v^3
-2\pi Tv
-\beta v^2
-2
\right)
}.
\label{eq:CP}
\end{equation}

\begin{figure}[t]
\centering
\includegraphics[width=0.48\textwidth]{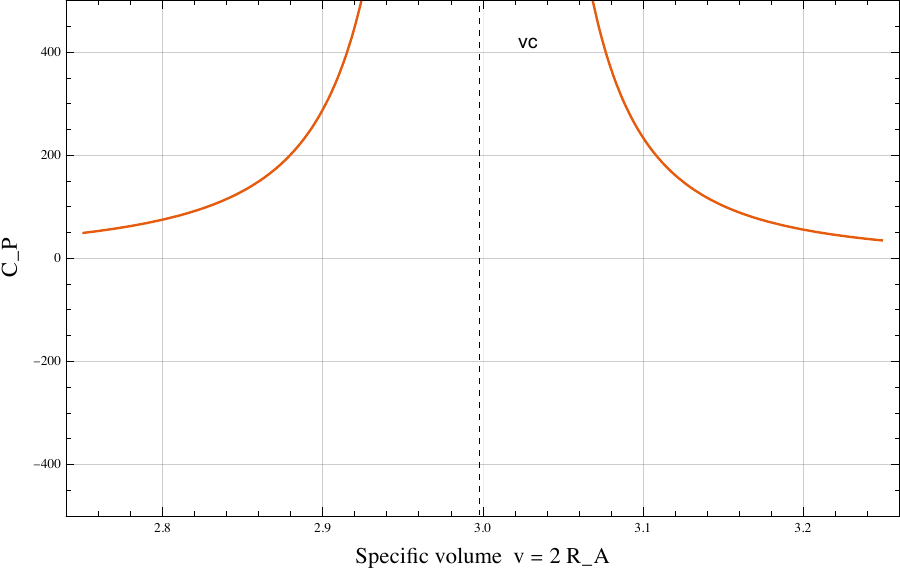}
\caption{Heat capacity at constant pressure $C_P$ evaluated along
the critical isotherm $T=T_c$. The singularity at $v=v_c$ follows
from the condition $(\partial P/\partial v)_T=0$, which coincides
with the spinodal locus.}
\label{fig:Cp}
\end{figure}

\begin{figure}[t]
\centering
\includegraphics[width=0.48\textwidth]{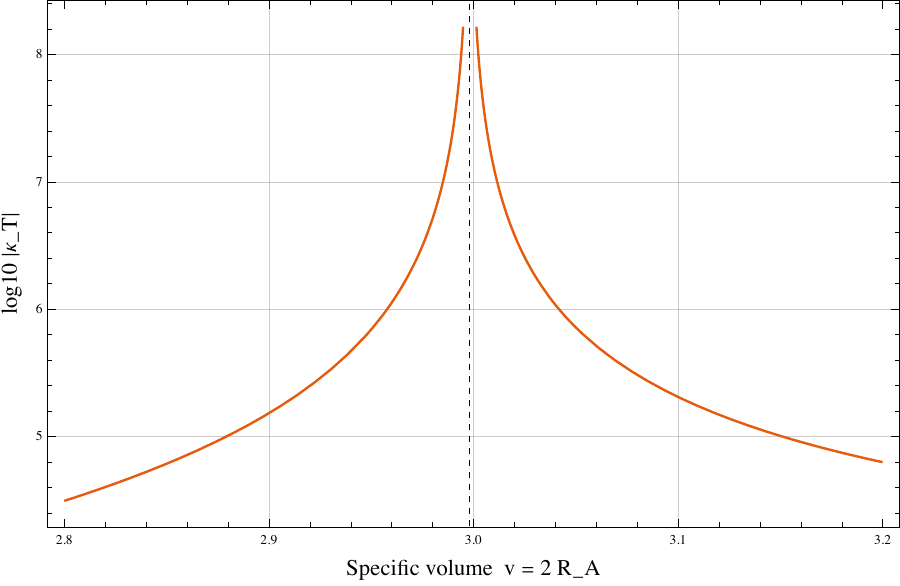}
\caption{Logarithmic representation of the isothermal
compressibility, $\log_{10}|\kappa_T|$, along the critical isotherm
$T=T_c$. The divergence at $v=v_c$ corresponds to the critical
point of the spinodal curve.}
\label{fig:kappa}
\end{figure}

It is useful to define
\begin{equation}
D(T,v)
=
\pi T\beta v^3
-2\pi Tv
-\beta v^2
-2.
\label{eq:Dresponse}
\end{equation}
Since
\begin{equation}
\left(\frac{\partial P}{\partial v}\right)_T
=
\frac{\Lambda e^x}{2\pi Gv^3}D(T,v),
\end{equation}
the heat capacity diverges whenever $D(T,v)=0$.
Now, the isothermal compressibility is defined by
\begin{equation}
\kappa_T
=
-\frac{1}{V}
\left(
\frac{\partial V}{\partial P}
\right)_T.
\label{eq:kappadef}
\end{equation}
Using $V=\pi v^3/6$, one obtains
\begin{equation}
\kappa_T
=
-\frac{3}{v}
\left[
\left(\frac{\partial P}{\partial v}\right)_T
\right]^{-1},
\end{equation}
and therefore
\begin{equation}
\kappa_T
=
-
\frac{
6\pi Gv^2e^{-x}
}{
\Lambda
\left(
\pi T\beta v^3
-2\pi Tv
-\beta v^2
-2
\right)
}.
\label{eq:kappaT}
\end{equation}
Equations~\eqref{eq:CP} and \eqref{eq:kappaT} show that both
$C_P$ and $\kappa_T$ possess the same singular locus,
\begin{equation}
D(T,v)=0
\qquad\Longleftrightarrow\qquad
\left(\frac{\partial P}{\partial v}\right)_T=0.
\label{eq:responsesing}
\end{equation}
This locus is precisely the spinodal curve discussed in the next
section. The critical point belongs to this curve and is
distinguished by the additional condition
\begin{equation}
\left(
\frac{\partial^2P}{\partial v^2}
\right)_T
=0.
\end{equation}
The sign of the isothermal compressibility provides the direct
criterion for local mechanical stability. Regions with $\kappa_T>0$ are mechanically stable, whereas $\kappa_T<0$ corresponds to a mechanically unstable branch. The divergences of
$C_P$ and $\kappa_T$ therefore identify the boundary separating the
different local thermodynamic branches. The divergence of the response functions along the spinodal curve is
the standard signature of the loss of local thermodynamic stability
\cite{Stanley1971}. Figures~\ref{fig:Cp} and \ref{fig:kappa} illustrate the behavior of
the response functions along the critical isotherm. Both quantities
become singular at $v=v_c$, consistently with the analytic
conditions above.

\section{Spinodal Curve}
\label{sec:spinodal}

\begin{figure}[t]
\centering
\includegraphics[width=0.48\textwidth]{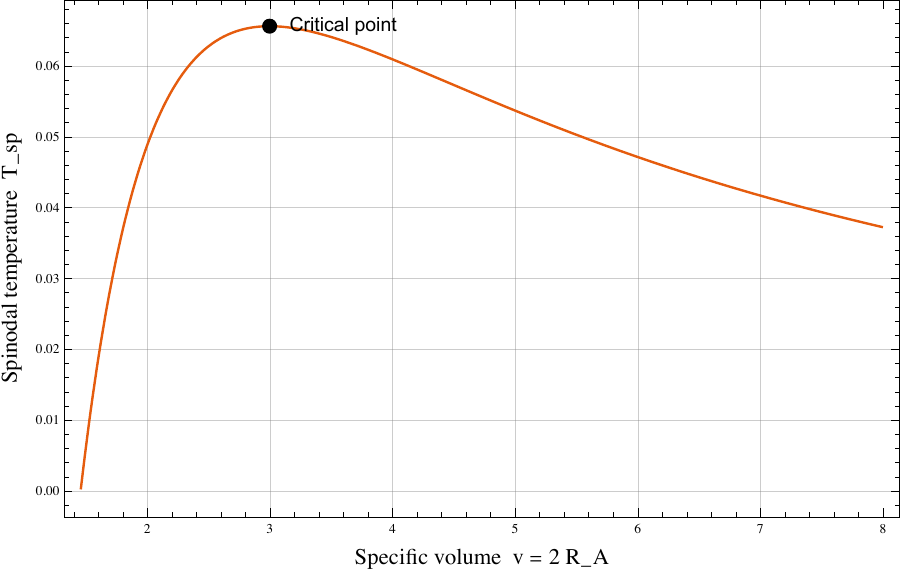}
\caption{Spinodal curve in the $(T,v)$ plane for the non-extensive
LQG-FLRW cosmological model. The curve is defined by
$(\partial P/\partial v)_T=0$, and its maximum coincides analytically
with the critical point $(v_c,T_c)$. For $T<T_c$, the two branches of
the spinodal delimit the mechanically unstable region in which
$\kappa_T<0$.}
\label{fig:spinodal}
\end{figure}

The spinodal curve determines the boundary of local mechanical
stability of the horizon system. Along this curve the isothermal
compressibility diverges, or equivalently,
\begin{equation}
\left(
\frac{\partial P}{\partial v}
\right)_T
=0.
\label{eq:spinodalcondition}
\end{equation}
Using Eq.~\eqref{eq:Pv}, the spinodal condition becomes
\begin{equation}
\pi T\beta v^3
-2\pi Tv
-\beta v^2
-2
=0.
\label{eq:spinodaleq}
\end{equation}
Solving for the temperature yields the exact spinodal curve
\begin{equation}
T_{\mathrm{sp}}(v)
=
\frac{
\beta v^2+2
}{
\pi v(\beta v^2-2)
}.
\label{eq:Tspinodal}
\end{equation}
The critical point corresponds to the extremum of the spinodal
temperature. Differentiating Eq.~\eqref{eq:Tspinodal} gives
\begin{equation}
\frac{dT_{\mathrm{sp}}}{dv}
=
-
\frac{
\beta^2v^4+8\beta v^2-4
}{
\pi v^2(\beta v^2-2)^2
}.
\label{eq:dTspinodal}
\end{equation}
Therefore,
\begin{equation}
\frac{dT_{\mathrm{sp}}}{dv}=0
\end{equation}
implies
\begin{equation}
\beta^2v^4+8\beta v^2-4=0.
\label{eq:spinodalcritical}
\end{equation}
This is exactly the same equation obtained from the simultaneous
criticality conditions
\begin{equation}
\left(
\frac{\partial P}{\partial v}
\right)_T
=
\left(
\frac{\partial^2P}{\partial v^2}
\right)_T
=0.
\end{equation}
Hence the extremum of the spinodal curve coincides analytically with
the thermodynamic critical point. For the physical branch,
\begin{equation}
\beta v_c^2
=
-4-2\sqrt5,
\end{equation}
and substitution into Eq.~\eqref{eq:Tspinodal} gives
\begin{equation}
T_{\mathrm{sp}}(v_c)
=
T_c
=
\frac{\sqrt5-1}{2\pi v_c}.
\label{eq:TspTc}
\end{equation}
The corresponding pressure along the spinodal is obtained
parametrically from
\begin{equation}
P_{\mathrm{sp}}(v)
=
P\left(T_{\mathrm{sp}}(v),v\right).
\label{eq:Pspinodal}
\end{equation}
Equations~\eqref{eq:Tspinodal} and \eqref{eq:Pspinodal} therefore
provide a parametric representation of the spinodal boundary in the
$(P,T)$ plane. For temperatures below the critical value, the equation
\begin{equation}
T=T_{\mathrm{sp}}(v)
\end{equation}
admits two solutions,
\begin{equation}
v_-(T)<v_+(T),
\end{equation}
which delimit the mechanically unstable branch. In this intermediate region $\kappa_T<0$, whereas the outer branches satisfy $\kappa_T>0$. The spinodal curve therefore characterizes local mechanical
stability. The distinction between globally stable and metastable
branches requires the Gibbs free-energy analysis presented in the
next section. Figure~\ref{fig:spinodal} shows the spinodal curve in the $(T,v)$
plane. Its maximum occurs at $(v_c,T_c)$, where the two spinodal
branches merge. For $T<T_c$, the two intersections of a horizontal
isotherm with the spinodal define the limits of the mechanically
unstable region.

Since
\begin{equation}
\kappa_T\propto D^{-1},
\end{equation}
whereas
\begin{equation}
R_N\propto-D^{-2},
\end{equation}
the leading critical behavior satisfies
\begin{equation}
|R_N|
\propto
\kappa_T^2.
\end{equation}
Using $\gamma_{cr}=1$, this is consistent with
\begin{equation}
R_N
\propto
-|t|^{-2\gamma_{cr}}
=
-|t|^{-2}.
\end{equation}

%=========================================================
\section{Gibbs Free Energy}
\label{sec:gibbs}
%=========================================================

The local stability analysis based on the response functions and the
spinodal curve does not determine which of the locally stable
thermodynamic branches is globally preferred. This information is
encoded in the Gibbs free energy. The thermodynamic construction employed below refers to the effective
state space of the apparent horizon, with the work density playing the
role of thermodynamic pressure, $P=W$. Accordingly, phase
coexistence should be understood as coexistence between distinct
horizon thermodynamic states, rather than as a microscopic material
phase transition of the cosmological fluid.

We begin with the unified first law of thermodynamics applied at the
apparent horizon,
\begin{equation}
T\,dS=-dE+W\,dV,
\label{eq:firstlaw_gibbs}
\end{equation}
where $E=\rho V$ is the total energy inside the horizon,
$W=(\rho-p)/2$ is the work density and $V=4\pi R_A^3/3$ is the volume
enclosed by the apparent horizon. Introducing the internal energy
$U=-E$ and identifying the thermodynamic pressure with the work
density, Eq.~\eqref{eq:firstlaw_gibbs} takes the standard
thermodynamic form $dU=T\,dS-P\,dV$. The identification $U=-E$
therefore follows directly from the horizon first law and does not
constitute an additional thermodynamic assumption.

\begin{figure}[t]
    \centering
    \includegraphics[width=0.85\columnwidth]
    {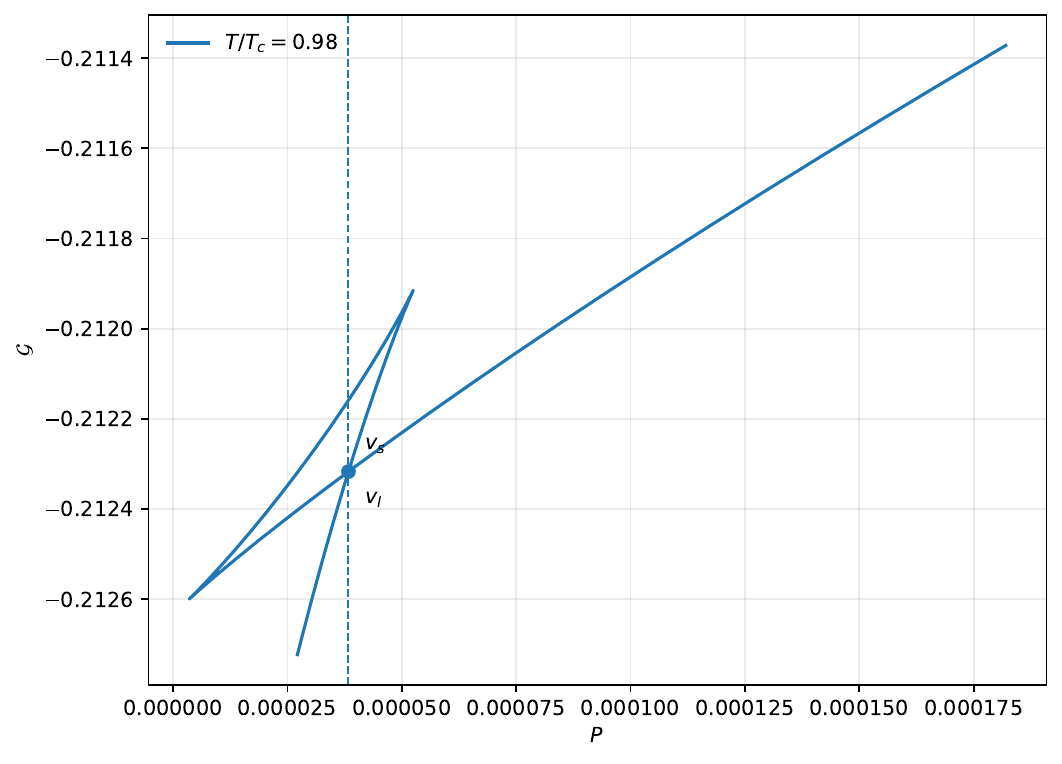}
    \caption{
Parametric Gibbs free energy $\mathcal{G}(P)$ at $T=0.98T_c$
for $q=1.3$ and $\Lambda=G=1$. The two marked states,
$v_s\simeq2.43101$ and $v_l\simeq3.97563$, have the same
coexistence pressure,
$P_{\mathrm{coex}}\simeq3.83492\times10^{-5}$,
and the same Gibbs free energy,
$\mathcal{G}_{\mathrm{coex}}\simeq-0.212317$.
They therefore represent distinct coexisting thermodynamic states
of the apparent horizon.
}
    \label{fig:gibbs}
\end{figure}

The Gibbs free energy is defined in the usual way and, using
$U=-\rho V$, can be written as
\begin{equation}
\mathcal{G}=U+PV-TS=(P-\rho)V-TS.
\label{eq:Gibbsdef}
\end{equation}
Using $P-\rho=-\tfrac{1}{2}(\rho+p)$ together with
Eq.~\eqref{eq:rhop}, we obtain
\begin{equation}
(P-\rho)V
=
\frac{\Lambda e^{\beta R_A^2}}{3G}
\left(
2\pi T R_A^2-R_A
\right).
\end{equation}
In terms of the specific volume $v=2R_A$ and the variable
$x=\beta v^2/4$ introduced in Sec.~\ref{sec:eos}, the Gibbs free
energy becomes
\begin{equation}
\mathcal{G}(T,v)
=
\frac{\Lambda e^x}{6G}
\left(
\pi Tv^2-v
\right)
-
\frac{T}{1-q}
\left(
e^x-1
\right).
\label{eq:Gibbs}
\end{equation}
As a consistency check, Eqs.~\eqref{eq:firstlaw_gibbs} and
\eqref{eq:Gibbsdef} imply
\begin{equation}
d\mathcal{G}
=
-S\,dT+V\,dP,
\label{eq:dGibbs}
\end{equation}
and therefore
\begin{equation}
\left(
\frac{\partial\mathcal{G}}{\partial P}
\right)_T
=
V
=
\frac{\pi v^3}{6}>0.
\label{eq:dGdP}
\end{equation}
This relation provides a direct thermodynamic consistency check for
the parametric Gibbs curves. At fixed temperature, the pressure and Gibbs free energy can be
regarded as parametric functions of the specific volume,
\begin{equation}
P=P(T,v),
\qquad
\mathcal{G}=\mathcal{G}(T,v).
\end{equation}
For $T<T_c$, the nonmonotonic equation of state produces multiple
branches in the $\mathcal{G}$-$P$ plane. The spinodal analysis
identifies which of these branches are locally stable or unstable,
while global coexistence requires two locally stable states,
$v_s$ and $v_l$, to satisfy simultaneously
\begin{equation}
P(T,v_s)
=
P(T,v_l)
=
P_{\mathrm{coex}},
\label{eq:Pequal}
\end{equation}
and
\begin{equation}
\mathcal{G}(T,v_s)
=
\mathcal{G}(T,v_l).
\label{eq:Gequal}
\end{equation}
The coexistence pressure may equivalently be determined through
Maxwell's equal-area construction
\cite{SpallucciSmailagic2013,KubiznakMann2012}.
\begin{equation}
\int_{V_s}^{V_l}
\left[
P(V,T)-P_{\mathrm{coex}}
\right]dV
=
0.
\label{eq:MaxwellV}
\end{equation}
Since
\begin{equation}
V=\frac{\pi v^3}{6},
\end{equation}
this condition becomes
\begin{equation}
\int_{v_s}^{v_l}
\left[
P(v,T)-P_{\mathrm{coex}}
\right]
\frac{\pi v^2}{2}\,dv
=
0.
\label{eq:Maxwellv}
\end{equation}
For the representative choice $q=1.3$ and $\Lambda=G=1$, and at $T=0.98T_c$, the coexistence conditions yield
\begin{equation}
v_s\simeq2.43101,
\qquad
v_l\simeq3.97563,
\end{equation}
with
\begin{equation}
P_{\mathrm{coex}}
\simeq
3.83492\times10^{-5},
\end{equation}
and
\begin{equation}
\mathcal{G}_{\mathrm{coex}}
\simeq
-0.212317.
\end{equation}
The two states therefore have the same temperature, pressure, and
Gibbs free energy, providing a direct thermodynamic verification of
phase coexistence between distinct horizon states. Figure~\ref{fig:gibbs} shows this coexistence explicitly. The two
marked points correspond to the states $v_s$ and $v_l$, which share
the same values of $P_{\mathrm{coex}}$ and
$\mathcal{G}_{\mathrm{coex}}$. As $T\rightarrow T_c$, the two
coexisting branches approach each other and merge at the critical
point.

\begin{figure}[t]
\centering
\includegraphics[width=0.85
\textwidth]{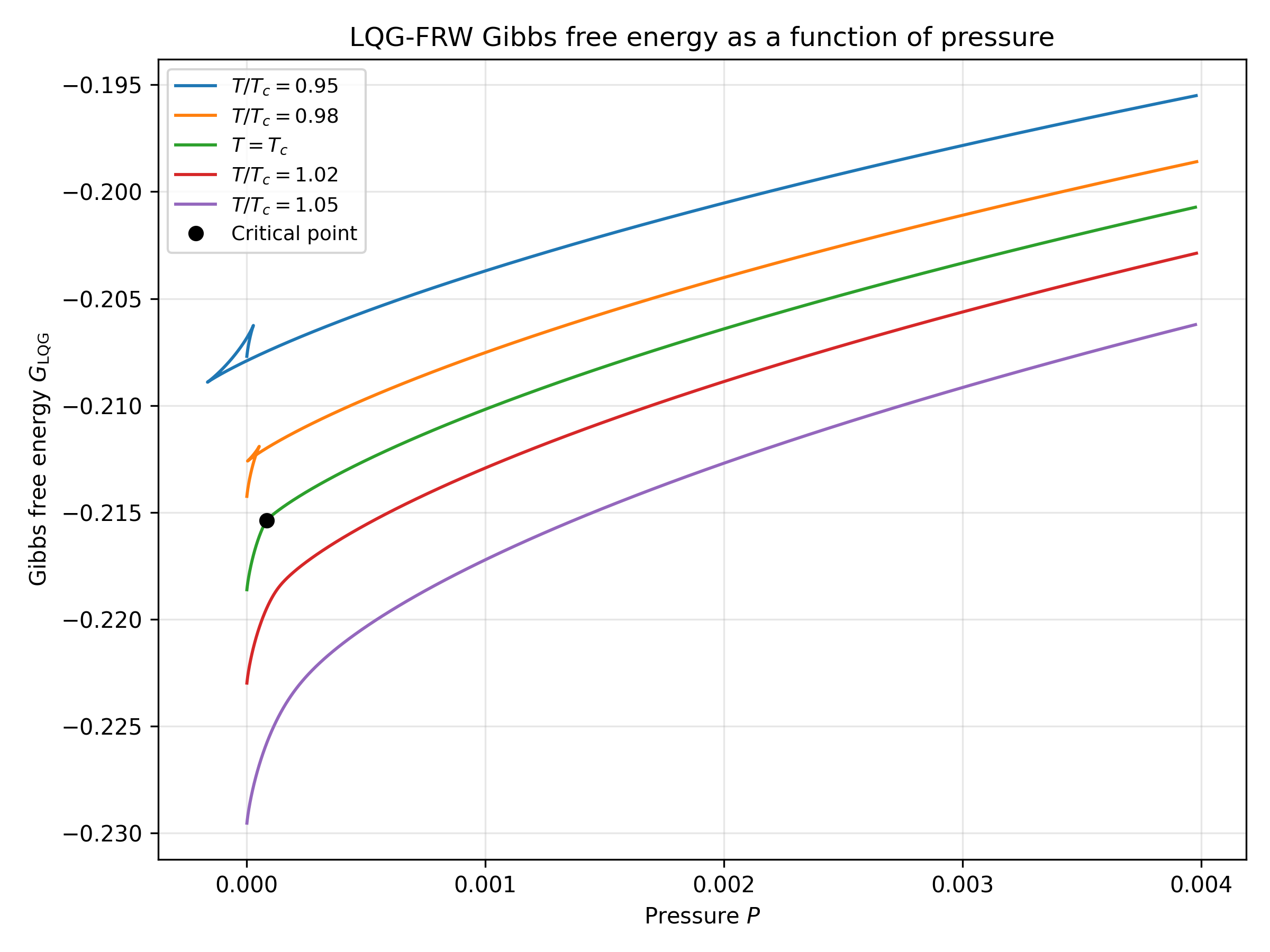}
\caption{Parametric Gibbs free energy $\mathcal{G}(P)$ for several
temperatures around the critical point. For $T<T_c$, the
nonmonotonic equation of state produces a multibranch structure
associated with locally stable and unstable thermodynamic states.
The branches merge as $T\rightarrow T_c$, while for $T>T_c$ the
curve becomes single-valued.}
\label{fig:Gibbs}
\end{figure}

\section{Critical Exponents and Universality Class}
\label{sec:critical_exponents}
We now determine the critical exponents governing the singular
behavior of the thermodynamic variables near the critical point.
The critical exponents are defined through the standard asymptotic
scaling relations near the critical point
\cite{Stanley1971,Goldenfeld1992}
\begin{equation}
t=\frac{T-T_c}{T_c},
\qquad
\phi=\frac{v-v_c}{v_c},
\qquad
p=\frac{P-P_c}{P_c}.
\label{eq:reduced_variables}
\end{equation}
Thus,
\begin{equation}
T=T_c(1+t),
\qquad
v=v_c(1+\phi).
\end{equation}
Expanding the equation of state around $(T_c,v_c)$ gives
\begin{equation}
p
=
a_{10}t
+
a_{11}t\phi
+
a_{03}\phi^3
+
O(t\phi^2,\phi^4,t^2),
\label{eq:landau_expansion}
\end{equation}
where
\begin{equation}
a_{10}
=
\frac{T_c}{P_c}
\left(
\frac{\partial P}{\partial T}
\right)_c,
\end{equation}
\begin{equation}
a_{11}
=
\frac{T_cv_c}{P_c}
\left(
\frac{\partial^2P}{\partial T\partial v}
\right)_c,
\end{equation}
and
\begin{equation}
a_{03}
=
\frac{v_c^3}{6P_c}
\left(
\frac{\partial^3P}{\partial v^3}
\right)_c.
\end{equation}
The absence of terms proportional to $\phi$ and $\phi^2$ follows
directly from the criticality conditions
\begin{equation}
\left(
\frac{\partial P}{\partial v}
\right)_c
=
0,
\qquad
\left(
\frac{\partial^2P}{\partial v^2}
\right)_c
=
0.
\end{equation}
For the physical critical branch, the coefficients $a_{11}$ and
$a_{03}$ are both finite and nonzero. For the representative choice
$q=1.3$ and $\Lambda=G=1$, one finds
\begin{equation}
a_{10}\simeq31.3287,
\qquad
a_{11}\simeq-164.039,
\qquad
a_{03}\simeq-61.1337.
\end{equation}
Thus, the critical point is an ordinary cubic critical point. The critical exponent $\alpha_{cr}$ characterizes the behavior of
the heat capacity at constant volume,
\begin{equation}
C_V\propto |t|^{-\alpha_{cr}}.
\end{equation}
Since the entropy depends only on the horizon radius, we found in
Sec.~\ref{sec:response} that $C_V=0$. Therefore, $\alpha_{cr}=0$.
The order parameter may be defined as the difference between the
large- and small-volume branches,
\begin{equation}
\eta=v_l-v_s.
\end{equation}
Although the physical thermodynamic volume is
\begin{equation}
V=\frac{\pi v^3}{6},
\end{equation}
near the critical point one has
\begin{equation}
V-V_c
=
\frac{\pi v_c^2}{2}(v-v_c)
+
O\left((v-v_c)^2\right),
\end{equation}
so that using $v_l-v_s$ or $V_l-V_s$ as the order parameter yields
the same critical exponent. For $t<0$, the coexistence conditions applied to
Eq.~\eqref{eq:landau_expansion} give, at leading order,
\begin{equation}
\phi_l=-\phi_s
\propto
(-t)^{1/2}.
\end{equation}
Consequently,
\begin{equation}
\eta
\propto
(-t)^{1/2},
\end{equation}
and therefore
\begin{equation}
\beta_{cr}=\frac{1}{2}.
\end{equation}
The exponent $\gamma_{cr}$ describes the divergence of the
isothermal compressibility,
\begin{equation}
\kappa_T
\propto
|t|^{-\gamma_{cr}}.
\end{equation}
Along the critical isochore $\phi=0$, Eq.~\eqref{eq:landau_expansion}
gives
\begin{equation}
\left(
\frac{\partial p}{\partial\phi}
\right)_t
=
a_{11}t+O(t^2).
\end{equation}
Since $a_{11}\neq0$, it follows that
\begin{equation}
\kappa_T
\propto
\frac{1}{|t|},
\end{equation}
and hence $\gamma_{cr}=1$.
Along the critical isotherm $t=0$, the leading-order equation of
state reduces to
\begin{equation}
p
=
a_{03}\phi^3
+
O(\phi^4).
\end{equation}
Since $a_{03}\neq0$,
\begin{equation}
|P-P_c|
\propto
|v-v_c|^3,
\end{equation}
and therefore $\delta_{cr}=3$.
The complete set of critical exponents is therefore
\begin{equation}
(\alpha_{cr},\beta_{cr},\gamma_{cr},\delta_{cr})
=
\left(
0,\frac{1}{2},1,3
\right).
\label{eq:critical_exponents}
\end{equation}
These are the standard mean-field critical exponents
\cite{Stanley1971,Goldenfeld1992}. They satisfy the standard scaling relations
\begin{equation}
\alpha_{cr}+2\beta_{cr}+\gamma_{cr}=2,
\end{equation}
\begin{equation}
\alpha_{cr}+\beta_{cr}(1+\delta_{cr})=2,
\end{equation}
and
\begin{equation}
\gamma_{cr}
=
\beta_{cr}(\delta_{cr}-1).
\end{equation}

Thus, despite the non-algebraic exponential-integral structure of
the equation of state, the critical point belongs to the standard
mean-field universality class. The non-extensive entropy deformation
determines the existence and thermodynamic location of the critical
point, while the local critical scaling remains of ordinary
mean-field type. This mean-field behavior is noteworthy because the present equation
of state is not of the algebraic Van der Waals form. Instead, it
contains both exponential and exponential-integral contributions
inherited from the non-extensive LQG-inspired entropy.

The origin of the mean-field behavior can be understood directly
from the local structure of the equation of state. At an ordinary
critical point,
\begin{equation}
\left(
\frac{\partial P}{\partial v}
\right)_c
=
\left(
\frac{\partial^2P}{\partial v^2}
\right)_c
=
0,
\end{equation}
while
\begin{equation}
\left(
\frac{\partial^2P}{\partial T\partial v}
\right)_c
\neq0,
\qquad
\left(
\frac{\partial^3P}{\partial v^3}
\right)_c
\neq0.
\end{equation}
Consequently, the leading nontrivial expansion around the critical
point has the generic form
\begin{equation}
p
=
a_{10}t
+
a_{11}t\phi
+
a_{03}\phi^3
+\cdots ,
\end{equation}
which is sufficient to generate the standard mean-field critical
exponents. This remark separates two conceptually different aspects of the thermodynamics. The entropy deformation determines the existence of a finite critical point and fixes its position in the thermodynamic parameter space. The local universality class is, however, determined by the analytic structure of the equation of state near that critical point. Hence the present non-extensive LQG-inspired model is an example of a system where ordinary mean-field critical scaling is preserved despite a strongly non-algebraic equation of state. In this sense the microscopic entropy deformation changes the global thermodynamic structure without changing the local critical exponents.
Similar mean-field behavior has been reported in several
gravitational thermodynamic systems with generalized entropy or
modified horizon dynamics
\cite{KubiznakMann2012,AbdusattarScalarTensor2023,
Housset2024,Rivadeneira2026}. However, whether this behavior persists
must be established model by model, since the existence of a critical
point and the required analyticity conditions are not guaranteed for
an arbitrary entropy functional.

\section{Cosmological Interpretation of the Critical Scale}
\label{sec:cosmological}
The thermodynamic criticality found in the previous sections is induced by the same entropy deformation which modifies the cosmological Friedmann equation. Thus, it is useful to understand how the thermodynamic critical scale relates to the rate of cosmological expansion. The modified Friedmann equation obtained from the non-extensive
LQG-inspired entropy is
\begin{equation}
\rho(H)
=
\frac{3\Lambda}{8\pi G}
\left[
H^2 e^{\beta/H^2}
-
\beta
\operatorname{Ei}
\left(
\frac{\beta}{H^2}
\right)
\right],
\label{eq:friedmann_cosmo}
\end{equation}
where
\begin{equation}
\beta
=
\frac{\pi(1-q)\Lambda}{G}.
\end{equation}
The physical critical branch requires $q>1$, and therefore $\beta<0$.
The departure from the standard Friedmann dynamics is controlled by
the dimensionless combination
\begin{equation}
x_H
=
\frac{\beta}{H^2}.
\label{eq:xH}
\end{equation}
Consequently, the characteristic scale associated with the entropy
deformation is
\begin{equation}
H_\beta
\sim
\sqrt{|\beta|}.
\label{eq:Hbeta}
\end{equation}
For
\begin{equation}
\frac{|\beta|}{H^2}\ll1,
\end{equation}
the entropy deformation is small and the dynamics approaches the
Bekenstein-Hawking regime. By contrast, when
\begin{equation}
H^2\sim|\beta|,
\end{equation}
the deformation becomes of order unity, and a perturbative expansion
around the Bekenstein-Hawking limit is no longer adequate.
The thermodynamic specific volume is related to the Hubble parameter
through
\begin{equation}
v
=
2R_A
=
\frac{2}{H}.
\label{eq:vH}
\end{equation}
Therefore, the critical specific volume determines a corresponding
critical expansion scale,
\begin{equation}
H_c
=
\frac{2}{v_c}.
\label{eq:Hcdef}
\end{equation}
An equivalent characterization follows directly from the
dimensionless deformation parameter $x_H=\beta/H^2$. At the critical
point,
\begin{equation}
x_{H,c}
=
\frac{\beta}{H_c^2}
=
-\frac{1}{2(\sqrt{5}-2)}
=
-\frac{2+\sqrt{5}}{2}
\simeq -2.118.
\end{equation}
Thus, the critical point lies outside the small-deformation regime
$|x_H|\ll1$. In particular, the phase transition cannot be captured
by a perturbative expansion around the Bekenstein-Hawking limit.
The critical behavior emerges only when the entropy deformation
becomes dynamically significant.
Using
\begin{equation}
v_c^2
=
\frac{-4-2\sqrt5}{\beta},
\end{equation}
and the physical branch $\beta<0$, we obtain
\begin{equation}
H_c^2
=
2(\sqrt5-2)|\beta|,
\label{eq:Hc}
\end{equation}
or equivalently
\begin{equation}
H_c
=
\sqrt{2(\sqrt5-2)}
\sqrt{|\beta|}
\simeq
0.6871\sqrt{|\beta|}.
\label{eq:Hc_numeric}
\end{equation}

The critical point therefore does not introduce an independent
cosmological scale. Instead, it occurs at an expansion rate set
directly by the entropy-deformation scale. This relation becomes particularly transparent by evaluating the
dimensionless deformation variable at the critical point. Since $x_H=\beta/H^2=\beta v^2/4$, one finds $x_c=\beta/H_c^2=-1-\sqrt5/2\simeq-2.11803$, and hence $\left|\beta/H_c^2\right|\simeq2.11803$. The critical point therefore lies outside the perturbative regime
$|\beta|/H^2\ll1$. The phase transition is intrinsically associated
with the nonperturbative sector of the entropy deformation. The extensive limit provides another useful interpretation. As
$q\rightarrow1$, one has $\beta\rightarrow0$, and consequently
\begin{equation}
v_c\rightarrow\infty,
\qquad
T_c\rightarrow0,
\qquad
H_c\rightarrow0.
\label{eq:GRcritical_limit}
\end{equation}
Hence the finite thermodynamic critical point is continuously removed in the Bekenstein-Hawking limit by being pushed to the degenerate infrared boundary of the thermodynamic state space. This is in accordance with the absence of finite criticality in the standard FLRW equation of state. However, it is essential to distinguish the existence of a thermodynamic critical point from the realization of one on a physical cosmological trajectory. A cosmological solution must fulfill the modified Friedmann equation, the continuity equation and the equation of state of the matter sector. Thus, existence of $(T_c,v_c,P_c)$ in the thermodynamic state space does not ensure the crossing of this point in the cosmological evolution. To determine if the critical regime can be reached in a realistic cosmological history one needs to specify the matter content and solve for $H(z)$. The present analysis determines the thermodynamic structure and identifies the corresponding expansion scale $H_c$; whether this critical regime is dynamically accessible is a separate cosmological question. Remarkably, $x_{H,c}$ is independent of the particular values of
$q$, $\Lambda$, and $G$; these parameters determine the physical
critical scale through $\beta$, while the reduced deformation at
criticality is fixed entirely by the criticality conditions.

\section{Thermodynamic Geometry and Ruppeiner Curvature}
\label{sec:ruppeiner}
Ruppeiner geometry provides a geometric formulation of equilibrium
thermodynamics in which the scalar curvature of the thermodynamic
state space can be used as a diagnostic of thermodynamic interactions
and critical behavior
\cite{Ruppeiner1979,Ruppeiner1995,Ruppeiner2010,Ruppeiner2012}.
Thermodynamic geometry provides an independent characterization of
the stability and critical structure obtained from the equation of
state. In the Ruppeiner formulation, the scalar curvature of the
thermodynamic state space is commonly interpreted as an effective
measure of thermodynamic interactions and correlations.
For the present horizon system $C_V=0$, and the standard two-dimensional Ruppeiner metric becomes degenerate. For systems with $C_V=0$, it is useful to introduce the normalized
thermodynamic curvature, which remains finite away from the
instability locus and retains the relevant information about the
critical behavior
\cite{WeiLiuMann2019,XuWuYang2020,WuWangXuYang2021}. We therefore consider the normalized Ruppeiner
curvature
\begin{equation}
R_N
=
C_V R_R,
\label{eq:RNdef}
\end{equation}
which remains finite in the formal $C_V\rightarrow0$ construction
and retains the singular structure of the thermodynamic geometry.
For an equation of state linear in the temperature, the normalized
curvature can be written as
\begin{equation}
R_N
=
\frac{1}{2}
-
\frac{T^2}{2}
\left[
\frac{\partial^2P/\partial T\,\partial v}
{\left(\partial P/\partial v\right)_T}
\right]^{2}.
\label{eq:RN_general}
\end{equation}

For the equation of state
\begin{equation}
P(T,v)
=
\frac{\Lambda}{G}e^x
\left(
\frac{T}{v}
+
\frac{1}{2\pi v^2}
\right)
-
\frac{3\Lambda\beta}{8\pi G}
\operatorname{Ei}(x),
\qquad
x=\frac{\beta v^2}{4},
\end{equation}
one obtains
\begin{equation}
\left(
\frac{\partial P}{\partial v}
\right)_T
=
\frac{\Lambda e^x}{2\pi Gv^3}
D(T,v),
\label{eq:Pv_RN}
\end{equation}
where
\begin{equation}
D(T,v)
=
\pi T\beta v^3
-
2\pi Tv
-
\beta v^2
-
2,
\label{eq:D_RN}
\end{equation}
and
\begin{equation}
\frac{\partial^2P}{\partial T\partial v}
=
\frac{\Lambda e^x}{2Gv^2}
\left(
\beta v^2-2
\right).
\label{eq:PTv_RN}
\end{equation}
Substitution into Eq.~\eqref{eq:RN_general} yields
\begin{equation}
R_N(T,v)
=
\frac{1}{2}
-
\frac{
\pi^2T^2v^2
\left(
\beta v^2-2
\right)^2
}{
2D(T,v)^2
}.
\label{eq:RN}
\end{equation}

The overall factor $\Lambda/G$ cancels identically. The normalized
curvature therefore depends on the thermodynamic variables and on
the entropy deformation through $\beta$, but not explicitly on the
overall normalization of the equation of state. Equation~\eqref{eq:Pv_RN} immediately implies
\begin{equation}
D(T,v)=0
\qquad\Longleftrightarrow\qquad
\left(
\frac{\partial P}{\partial v}
\right)_T=0.
\label{eq:RN_spinodal}
\end{equation}
Hence, provided that the numerator of Eq.~\eqref{eq:RN} does not
vanish simultaneously,
\begin{equation}
D(T,v)\rightarrow0
\qquad\Longrightarrow\qquad
R_N\rightarrow-\infty.
\label{eq:RNdiv}
\end{equation}
The singular locus of the thermodynamic curvature therefore
coincides exactly with the spinodal curve. This provides an
independent geometric characterization of the mechanical
instability boundary. The critical point is distinguished from a generic point of the
spinodal by the additional condition
\begin{equation}
\left(
\frac{\partial^2P}{\partial v^2}
\right)_c=0.
\end{equation}
Since
\begin{equation}
\left(
\frac{\partial P}{\partial v}
\right)_T
=
A(v)D(T,v),
\qquad
A(v)
=
\frac{\Lambda e^x}{2\pi Gv^3},
\end{equation}
the criticality conditions imply
\begin{equation}
D(T_c,v_c)=0,
\qquad
\left.
\frac{\partial D}{\partial v}
\right|_c
=0.
\end{equation}
Consequently, along the critical isotherm,
\begin{equation}
D(T_c,v)
=
\frac{1}{2}
\left.
\frac{\partial^2 D}{\partial v^2}
\right|_c
(v-v_c)^2
+\cdots .
\end{equation}
Using
\begin{equation}
\left.
\frac{\partial^2 D}{\partial v^2}
\right|_c
=
-\frac{2(5+\sqrt{5})}{v_c^2},
\end{equation}
one obtains
\begin{equation}
D(T_c,v)
\simeq
-\frac{5+\sqrt{5}}{v_c^2}
(v-v_c)^2.
\end{equation}
Moreover, the numerator of Eq.~\eqref{eq:RN} remains finite and
nonzero at the critical point. In fact,
\begin{equation}
\frac{\pi^2 T_c^2 v_c^2}{2}
\left(
\beta v_c^2-2
\right)^2
=
4(3+\sqrt{5}).
\end{equation}
Therefore, the leading critical behavior is
\begin{equation}
R_N(T_c,v)
\simeq
-\frac{2}{5}
\left(
\frac{v_c}{v-v_c}
\right)^4.
\label{eq:RN_v_scaling}
\end{equation}
Equivalently, in terms of the reduced volume variable
\begin{equation}
\phi
=
\frac{v-v_c}{v_c},
\end{equation}
the critical divergence takes the compact form
\begin{equation}
R_N(T_c,v)
\simeq
-\frac{2}{5}\phi^{-4}.
\end{equation}
A complementary scaling law follows along the critical isochore.
Since $D(T,v)$ is linear in $T$,
\begin{equation}
D(T,v_c)
=
\left.
\frac{\partial D}{\partial T}
\right|_c
(T-T_c),
\end{equation}
with
\begin{equation}
\left.
\frac{\partial D}{\partial T}
\right|_c
=
\pi v_c
\left(
\beta v_c^2-2
\right)
=
-2\pi v_c(3+\sqrt{5}),
\end{equation}
one obtains
\begin{equation}
R_N(T,v_c)
\simeq
-\frac{1}{
\pi^2 v_c^2(3+\sqrt{5})
}
\frac{1}{(T-T_c)^2}.
\end{equation}
Introducing the reduced temperature
\begin{equation}
t
=
\frac{T-T_c}{T_c},
\end{equation}
and using
\begin{equation}
T_c
=
\frac{\sqrt{5}-1}{2\pi v_c},
\end{equation}
the previous expression simplifies to
\begin{equation}
R_N(T,v_c)
\simeq
-\frac{1}{2t^2}.
\label{eq:RN_t_scaling}
\end{equation}
This result is directly related to the divergence of the isothermal
compressibility. Since
\begin{equation}
\kappa_T\propto D^{-1},
\end{equation}
whereas
\begin{equation}
R_N\propto-D^{-2},
\end{equation}
the leading critical behavior satisfies
\begin{equation}
|R_N|
\propto
\kappa_T^2.
\label{eq:RN_kappa}
\end{equation}
Using $\gamma_{cr}=1$, this gives
\begin{equation}
R_N
\propto
-|t|^{-2\gamma_{cr}}
=
-|t|^{-2}.
\end{equation}

\begin{figure}[t]
    \centering
    \includegraphics[width=0.72\linewidth]{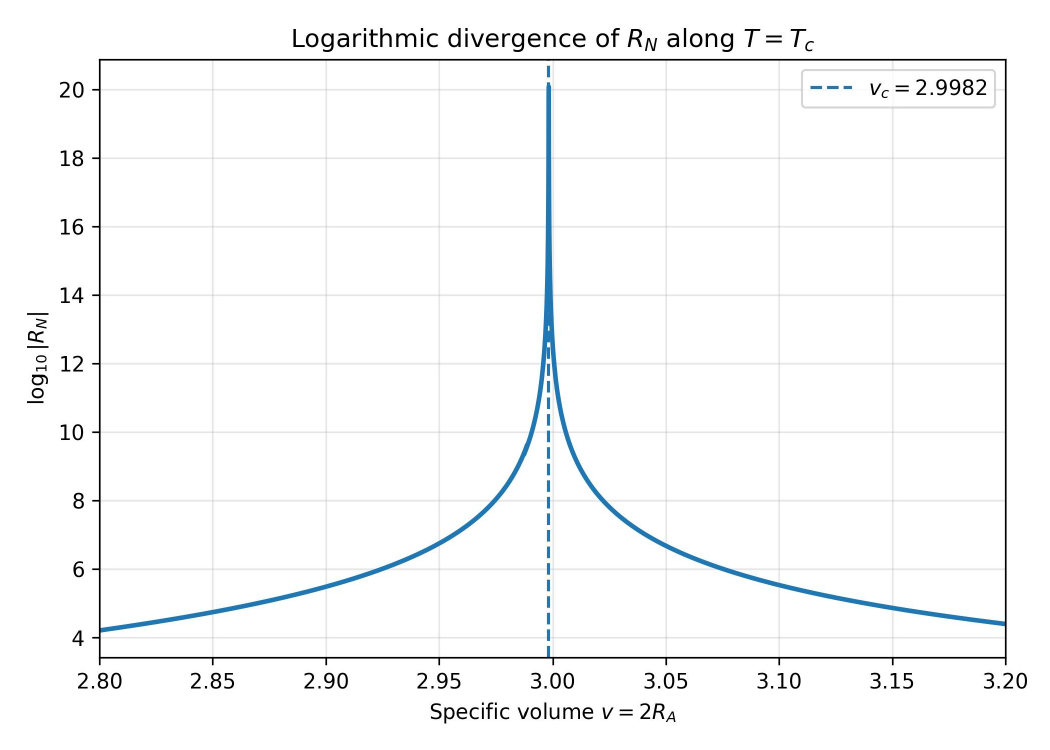}
    \caption{Behavior of $\log_{10}|R_N|$ along the critical isotherm
    $T=T_c$. The divergence at $v=v_c$ corresponds to the critical
    point where the two branches of the spinodal curve merge.}
    \label{fig:ruppeiner_critical}
\end{figure}

The normalized curvature is mostly negative away from the singular locus in the thermodynamically stable region.
Within the standard Ruppeiner interpretation, negative thermodynamic
curvature is commonly associated with effectively attractive
interactions
\cite{Ruppeiner1995,Ruppeiner2010}.
Such an interpretation should be considered as a macroscopic thermodynamic diagnostic, rather than as a microscopic identification of the underlying quantum-gravitational degrees of freedom.
Thus, the thermodynamic geometry gives the same stability boundary as obtained independently from the equation of state, response functions and spinodal analysis. More importantly, the critical divergence of $R_N$ obeys definite scaling laws and is directly related to the divergence of the mechanical susceptibility.
\begin{figure}[t]
    \centering
    \includegraphics[width=0.78\linewidth]{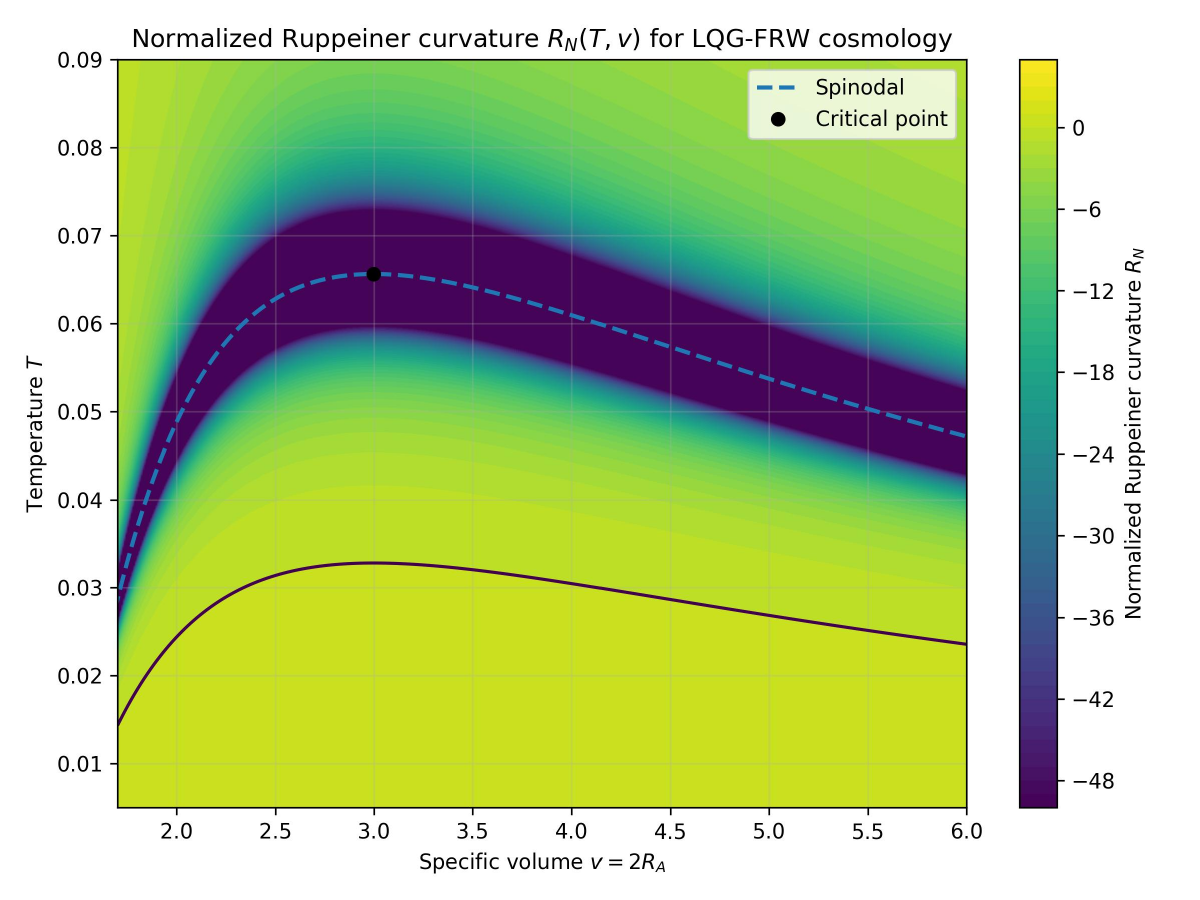}
    \caption{Contour plot of the normalized Ruppeiner curvature
    $R_N(T,v)$ for the LQG-FLRW cosmological model. The dashed curve
    denotes the spinodal line defined by
    $(\partial P/\partial v)_T=0$, while the black dot marks the
    critical point $(v_c,T_c)$. The singular structure of $R_N$
    follows the spinodal boundary and the curvature is predominantly
    negative in the thermodynamically stable region.}
    \label{fig:ruppeiner_contour}
\end{figure}
A complementary view is provided by
Fig.~\ref{fig:ruppeiner_critical}, where
$\log_{10}|R_N|$ is evaluated along the critical isotherm $T=T_c$.
The sharp divergence at $v=v_c$ provides a direct numerical
verification of the analytical singularity predicted by
Eq.~\eqref{eq:RN}.

The thermodynamic geometry therefore gives an alternative description of the phase structure obtained from the equation of state, response functions, spinodal curve and Gibbs free energy. In particular the divergence of $R_N$ reproduces the mechanical instability boundary and its singular behavior at the critical point is consistent with the development of long-range correlations.
Together with the mean-field critical exponents obtained in the last section, these results indicate that the LQG-corrected apparent horizon exhibits the usual thermodynamic signatures of an interacting critical system.

%\begin{figure}[t]
%\centering
%\includegraphics[width=0.48\textwidth]{LQG_FRW_Ruppeiner_contour.pdf}
%\caption{Contour plot of the normalized Ruppeiner curvature
%$R_N(T,v)$ for the LQG-FLRW cosmological model. The dashed 5
%curve
%denotes the spinodal line and the black dot marks the %critical point
%$(v_c,T_c)$. The divergence of $R_N$ along the spinodal %provides a
%geometric characterization of the mechanical-instability %boundary.}
%\label{fig:RuppeinerContour}
%\end{figure}
%\begin{figure}[t]
%\centering
%\includegraphics[width=0.48\textwidth]{LQG_FRW_Ruppeiner_RN_log_Tc.pdf}
%\caption{Logarithmic representation of the normalized %Ruppeiner
%curvature, $\log_{10}|R_N|$, along the critical isotherm %$T=T_c$.
%The divergence at $v=v_c$ agrees with the analytical critical
%scaling of the thermodynamic curvature.}
%\label{fig:RuppeinerCritical}
%\end{figure}
Thermodynamic geometry has also been applied to the phase structure
of FLRW apparent horizons, providing a complementary characterization
of cosmological criticality
\cite{AbdusattarRuppeiner2023,Housset2024}. Future work could involve confronting the modified Friedmann equations with observational data, extending the analysis to non-flat or anisotropic cosmologies, other quantum-gravity-inspired entropy formalisms, and possible connections between thermodynamic phases and the ultimate fate of the universe, such as the de Sitter, Big Rip and Little Rip scenarios.

\section{Conclusions}
\label{sec:conclusions}
In this work we have investigated the thermodynamic critical behavior
of a spatially flat FLRW universe whose apparent horizon is endowed
with a non-extensive LQG-inspired entropy. Starting from the horizon
first law and retaining the full Kodama-Hayward temperature, we
derived the modified Friedmann dynamics and constructed the
corresponding horizon equation of state.
The entropy deformation generates a finite critical point only in
the non-extensive branch $q>1$. In the Bekenstein-Hawking limit
$q\to1$, the critical volume is pushed to infinity while the
critical temperature and the corresponding expansion scale vanish.
This shows that the finite critical structure is induced by the
entropy deformation rather than inherited from standard FLRW horizon
thermodynamics.
The local and global thermodynamic analyses provide a mutually
consistent picture. The heat capacity at constant pressure and the
isothermal compressibility diverge along the spinodal curve, whose
extremum coincides with the critical point. Below the critical
temperature, the Gibbs free energy develops multiple branches, and
the coexistence of distinct horizon thermodynamic states is verified
by requiring equality of temperature, pressure, and Gibbs free
energy. This coexistence refers to the effective thermodynamic state
space of the apparent horizon and should not be interpreted as a
microscopic material phase transition of the cosmological fluid.
The critical exponents take the standard mean-field values. Thus,
although the entropy deformation determines the existence and
location of the critical point, the local critical behavior remains
in the mean-field universality class. This separation between the
global thermodynamic effects of the entropy deformation and the
local universality of the critical point is one of the main results
of the present analysis.
The normalized Ruppeiner curvature provides an independent geometric
characterization of the same structure. Its singular locus coincides
with the spinodal curve, and its critical divergence is directly
related to the divergence of the isothermal compressibility. The
predominantly negative curvature in the stable region is consistent,
within the usual Ruppeiner interpretation, with effectively
attractive thermodynamic interactions, without implying a specific
microscopic identification of the underlying quantum-gravitational
degrees of freedom.
The critical point also defines a characteristic cosmological
expansion scale controlled by the same entropy-deformation parameter
that modifies the Friedmann equation. The critical regime lies
outside the small-deformation limit, indicating that the phase
transition cannot be captured as a perturbative correction around
the Bekenstein-Hawking regime.
It is important, however, to distinguish thermodynamic criticality
in the horizon state space from its realization along a physical
cosmological trajectory. The existence of a critical point does not
imply that an evolving universe necessarily reaches or crosses it.
Addressing this question requires specifying the matter sector and
solving the modified cosmological dynamics. Determining whether the
critical regime is dynamically accessible and whether it can leave
observable signatures in the expansion or perturbation histories
constitutes a natural direction for future work.

\bibliography{biblio.bib}
\bibliographystyle{elsarticle-num}

\end{document}